\documentclass[universe,review,accept,moreauthors,pdftex]{Definitions/mdpi} 

\usepackage{color}

\firstpage{1} 
\pubvolume{1}
\issuenum{1}
\articlenumber{0}
\pubyear{2021}
\copyrightyear{2020}
\datereceived{} 
\dateaccepted{} 
\datepublished{} 
\hreflink{https://doi.org/} 

\Title{The zoo of isolated neutron stars}

\TitleCitation{The zoo of isolated neutron stars}

\Author{Sergei Popov $^{1}$*\orcidA{}}

\AuthorNames{Sergei Popov}

\AuthorCitation{Popov, S.}

\address{%
$^{1}$ \quad ICTP -- Abdus Salam International Center for Theoretical Physics, Strada Costiera 11, I-34151 Trieste, Italy}

\corres{Correspondence: sergepolar@gmail.com; Tel.:  +39 040 2240 368}

\abstract{In this brief review I summarize our basic knowledge about different types of isolated  neutron stars. I discuss radio pulsars, central compact objects in supernova remnants, magnetars, near-by cooling neutron stars (aka the Magnificent seven), and sources of fast radio bursts. Several scenarios of magneto-rotational evolution are presented. Recent observational data, in the first place -- discovery of long period radio pulsar, require non-trivial evolution of magnetics fields or/and spin periods of neutron stars. In some detail I discuss different models of magnetic field decay and interaction of young neutron stars with fallback matter. }

\keyword{neutron stars; radio pulsars; magnetars; fast radio bursts} 

\begin{document}

\section{Introduction}

Neutron stars (NSs) are very fascinating objects. More we study them -- more interesting they seem to be. 

There are numerous types of sources related to NSs. Many of them contain a NS as a member of a binary system. In the first place, these are X-ray binaries with accretion onto the compact object. The first discovery (but not identification!) of such source -- Sco X-1, -- has happened already in 1962 \cite{1964Natur.204..981G}. The first robust identification of a NS in an accreting binary -- Cen X-3, -- was done a few years later thanks to observations on-board of UHURU satellite  \cite{1971ApJ...167L..67G}.

Even as accreting objects NSs in close binary systems can appear as sources with very different properties. This is possible because binary evolution provides many possibilities to obtain systems with various parameters \cite{2023arXiv230308997B}.

In low-mass X-ray binaries (LMXBs) mass transfer from a low-mass component spins up the NS. This can result in formation of a millisecond radio pulsar (mPSR) \cite{1976SvAL....2..130B}. The first mPSR was discovered in 1982 \cite{1982Natur.300..615B}. Spin up in a binary system was confirmed after identification of so-called transitional mPSRs \cite{2013Natur.501..517P}.  Such NSs are characterized by very short spin periods ($\sim$1-10 msec) and low magnetic fields ($\sim 10^8$--$10^{10}$~G). The later leads to large characteristic ages, $\tau_\mathrm{ch}=P/(2\dot P)\sim$~few billion years, and long life time. 

Some of PSRs (ordinary and mPSRs) are found in binary systems with other NSs \cite{2019ApJ...876...18F}.
There are several evolutionary channels which can result in formation of such systems \cite{2017ApJ...846..170T}. Some of these binary NSs are doomed to end there lives in a spectacular coalescence accompanied by a short gamma-ray burst (sGRB), a gravitational wave burst, and a kilonova \cite{2017RPPh...80i6901B}.

NSs are usually found in binary systems due to their own activity (as in the case of PSRs), or due to their interaction with the companion (accretion or coalescence). 
However, recently several candidates to inactive/non-interacting NSs in binaries have been reported on the base of astrometric and spectroscopic data,
see \cite{2022arXiv220700680A} and references therein to other proposed candidates.

Observations of binary systems provide plethora of data on NSs. Still, in many cases it is necessary to study isolated NSs, as in this case properties of compact objects are not modified by presence of a companion. In this review I focus on different types of isolated (mostly -- young) NSs and their evolution. 

\section{Main species in the Zoo}

 Observations in the whole range of electromagnetic waves -- from radio to gamma, -- demonstrate very different phenomena related to isolated NSs. Various observational appearances and intrinsic properties (spin period, magnetic field, surface temperature, etc.) result in classification of isolated NSs into several main types, see Fig.~1. In this section I briefly present descriptions of them.

\subsection{Radio pulsars}

Radio pulsars (PSRs) are the best known (and the most numerous if we speak about observed objects) type of isolated NSs. 
They were discovered 55 years ago \cite{1968Natur.217..709H} by detection of their periodic radio pulses. The periodicity is due to rotation of these compact objects.  Spin periods cover the range $P\sim0.001$~--$100$~s. Presently, the ATNF catalogue \footnote{https:www.atnf.csiro.au/people/pulsar/psrcat/.} 
contains $\sim3000$ of these sources \cite{2005AJ....129.1993M}.
Magnetic fields determined from the standard spin-down equation $B_\mathrm{p}=3.2\times 10^{19} \left(P\dot P\right)^{1/2}$
are in the range $\sim10^8$~--~$10^{14}$~G. These are dipolar fields on magnetic poles. On the equator the field it twice smaller: $B=B_\mathrm{p}/2$. Non-dipolar components might also exist, but their determination in the case of radio pulsars is not possible with any confidence, at least now.

Low fields together with small spin periods correspond to old ``recycled'' mPSRs originated from LMXBs. In this section we do not discuss them, focusing on young ``normal'' PSRs (in many respects, a group of so-called Rotating Radio Transients, -- RRATs, see \cite{2006Natur.439..817M} -- can be unified with PSRs as they share the main properties).

 Radio emission of PSRs is generated in magnetospheric processes which are not completely understood, yet (see a review and references to early studies in \cite{2018PhyU...61..353B}). 
Magnetospheric emission might have a very wide spectrum. In some cases, e.g. the Crab pulsar, it is observed in the whole spectral range: from radio to gamma-rays. 

The energy reservoir is related to rotation of a PSR. This allows to relate key properties of a NS with observed parameters in a simplified magneto-dipole formula:

\begin{equation}
 I \omega \dot \omega = \frac{2 \mu^2 \omega^4}{3 c^3}.
\end{equation}
Here $I$ is the moment of inertia of a NS, $\omega=2\pi/P$ -- spin frequency, $\mu = BR_\mathrm{NS}^3$ -- magnetic moment, and $c$ -- the speed of light. $R_\mathrm{NS}$ is the NS radius.
Detailed calculations broadly confirm this equation \cite{2014MNRAS.441.1879P}, still alternative views also exist e.g., \cite{2019MNRAS.485.4573P}.

Different studies indicate that the majority of young NSs passes through the stage of a PSR. 
Thus, the birth rate of PSRs is not much smaller than the birth rate of NSs in general. The later, in its turn, is not much smaller than the rate of core-collapse supernovae (CCSN) which is about 1/60~yr$^{-1}$ \cite{2021NewA...8301498R}. 
All these numbers are known with an uncertainty smaller than a factor $\sim 2$. Let us assume for a simple estimate that the birth rate of PSRs is 1/100 yr$^{-1}$. Near the so-called ``death line'', where the number of observed pulsars in the $P$~--~$\dot P$ diagram drops (this might be related to a drop in efficiency of $ee^-$-pair production in the magnetosphere), a typical pulsar has $P\sim2$~s and $\dot P \sim 3\times 10^{-16}$~s/s. Then its  characteristic age is about $10^8$~yrs. So, the total number of PSRs in the Galaxy is $\sim  10^6$. As radio emission of PSRs is significantly beamed, just a small fraction (about 10\%, \cite{1998MNRAS.298..625T}) of them can be observed even with infinite sensitivity. At the moment, the Five-hundred-meter Aperture Spherical radio Telescope (FAST) is the most sensitive instrument looking for new PSRs \cite{2021RAA....21..107H, 2021ApJ...915L..28P}. 
It is expected that SKA will allow us to detect most of the Galactic pulsars potentially visible from the Earth \cite{2018IAUS..337..171L}. 

PSRs are characterised by large spatial velocities, $\sim$ few hundred km~s$^{-1}$ -- about an order of magnitude larger than in the case of their progenitors \cite{1994Natur.369..127L}.  
Additional velocity, kick, is received by a NS during the SN explosion \cite{2013A&A...552A.126W}. For young NSs spatial velocity is not changing significantly. So, there were hopes that properties of velocity distributions of different subpopulations of isolated NSs could shed light on their origin and causes of diversity of their properties. However, it seems that velocities of different types of isolated NSs are similar to each other.

Young ages of normal PSRs are confirmed by their associations with supernova remnants (SNRs) \cite{2022MNRAS.514.4606I}
and spiral arms \cite{2021MNRAS.508.1929C}. 
Some types of isolated NSs that we are going to discuss are even younger on average and are associated with SNRs by definition. 

\begin{figure}[H]
\includegraphics[width=10.5 cm]{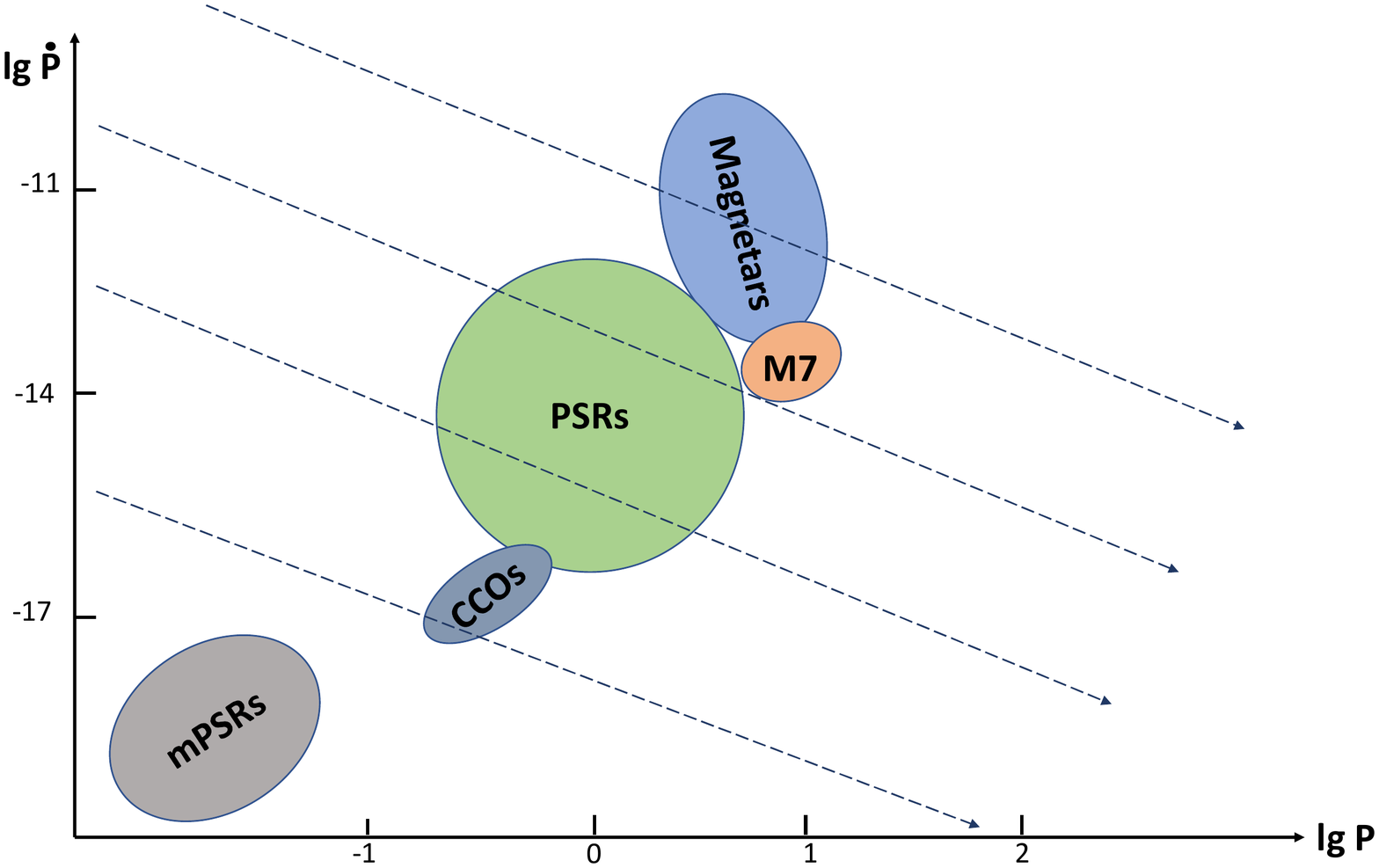}
\caption{$P$~--~$\dot P$ diagram. The main types of young isolated NSs are shown together with recycled mPSRs (lower left part of the plot) and the simplest tracks corresponding to Eq.~(1) with constant magnetic field. Scales on axes are approximate. \label{fig1}}
\end{figure}

\subsection{Central compact objects}

 Central compact objects in SNRs (CCOs for short) are young NSs observed due to their thermal emission inside supernova remnants. 
Known CCOs are not numerous, there about a dozen of them
\footnote{See the on-line catalogue at http://www.iasf-milano.inaf.it/$\sim$deluca/cco/main.htm. Mostly, parameters of CCOs mentioned in this subsection refer to this catalogue.} \cite{2017JPhCS.932a2006D, 2020MNRAS.495.1692G}. 
But their very small ages -- about few thousand years, -- point to a significant birth rate.
So, this is a non-negligible subpopulation of isolated NSs. 

CCOs can be an inhomogeneous group of sources united by their appearance inside SNRs and absence of any traces of radio pulsar activity. These objects are observed as soft X-ray sources with a thermal spectrum \cite{2000ApJ...531L..53P}. The origin of emission is attributed to hot areas of a NS surface with a typical temperature $\sim 10^6$~K. As sources are young (ages $\sim$~few thousand years), the most obvious source of energy is the residual heat. This assumption is in correspondence with modeling of thermal evolution of NSs \cite{2015SSRv..191..171P,2018A&A...609A..74P}. 
However, at least in few cases an additional source of energy is suspected. 

Standard cooling is determined by neutrino emission from interiors and photon emission from the surface \cite{2004ApJS..155..623P}. At early stage of evolution, typically up to $10^4$~--~$10^5$~yrs, the neutrino emission prevails. The neutrino emissivity strongly depends on the NS mass: low-mass objects can stay hot for a longer time. Still, residual heat can explain only temperatures $\lesssim 3 \times 10^6$~K for ages $\lesssim 10^5$~~yrs and $\lesssim  10^6$~K for ages $\lesssim 10^6$~~yrs. Larger temperatures at young ages are typically explained by magnetic energy release, e.g. \cite{2008ApJ...673L.167A}. Measurable temperature in mature NSs (e.g., \cite{2022ApJ...924..128A}) can be explained by chemical \cite{2021MNRAS.508.6118K} or rotochemical heating \cite{2020MNRAS.492.5508Y}.

 In three CCOs spin periods and their derivatives are measured due to the X-ray flux variability. 
 Periods $\sim 0.1$~--~0.4~s and $\dot P\sim 10^{-17}$~s/s according to the magneto-dipole formula correspond to fields $\sim 10^{10}$~--~$10^{11}$~G. Because of their position in the $P$~--~$\dot P$ diagram these CCOs are often called ``antimagnetars''. Still, some of CCOs can be real magnetars.

Analysis of the light curve of PSR J1852+0040 in the SNR Kes 79 allowed  Shabaltas \& Lai
\cite{2012ApJ...748..148S} to state that large pulse fraction observed in this source requires crustal field of the magnetar scale (see also \cite{2014ApJ...790...94B}). Additional energy release due to the field decay in the crust, or modification of the surface temperature distribution due to the influence of the magnetic field on the heat transfer might be responsible for the small emitting area in PSR J1852+0040 which is necessary to explain large pulsations of the flux in  presence of the light bending effect in strong gravitational field of the compact object. As no magnetospheric activity was ever observed from this source, it was proposed in \cite{2012ApJ...748..148S} that the NS is a ``hidden'' magnetar. I.e., the strong field is ``screened'' by the matter which fallback onto the NS soon after the SN explosion (see \cite{2013ApJ...770..106B} and references therein to early papers on this scenario). 

Another magnetar among CCOs is the famous NS inside the RCW 103 remnant. 
This source has been discovered years ago with the Einstein observatory \cite{1980ApJ...239L.107T}.
A prominent feature of the central source -- its variability on a time scale about a few years, -- was successfully described in the model of magnetic energy release in the crust \cite{2015PASA...32...18P}, and it was suspected that the source can also belong to the class of ``hidden'' magnetars. However, 
soon it was demonstrated that it is not so hidden. 
Bright short high energy bursts, quite similar to the bursts of soft gamma-ray repeaters, were detected from this source \cite{2016MNRAS.463.2394D, 2016ApJ...828L..13R}. This clearly points towards the magnetar nature of the source. In addition, the 6.67 hour spin period was measured for the NS in RCW103 \cite{2006Sci...313..814D}.  Such long spin periods in young presently non-accreting objects can be explained in a model where a strong magnetic field of the NS interacts with a fallback disc \cite{2022ApJ...934..184R}. 

 Despite the majority of CCOs demonstrate just surface thermal emission, at least in few cases CCOs can have an additional source of energy -- its magnetic field. I.e., they can be related to magnetars.

\subsection{Magnetars}

A NS is called a magnetar if its observational appearance is mainly due to magnetic energy release. 
There are two main manifestations of these sources: high energy bursts and surface thermal emission.
 Often magnetars are considered as the most extreme and interesting type of NSs. 
Their unusual properties manifest themselves in spectacular observational appearance, unfamiliar to other types of compact objects.
 Indeed, during the hyper flare of SGR 1806-20 in 2004 the peak luminosity was above $10^{47}$~erg~s$^{-1}$ and the total energy release in the event was $\sim 10^{46}$~erg \cite{2005Natur.434.1107P}. This tremendous burst demonstrates that magnetic energy in NSs can reach very large values and also can be rapidly released in a huge amount.

 It is not easy to say exactly when magnetars were discovered. May be, the most reasonable approach is to attribute it to the identification of the source which was called in 1979 as ``X-ray burster 0525.9-66.1'' \cite{1979Natur.282..587M, 1979PAZh....5..588V}, and which is now mostly known as SGR 0525-66. Observations with gamma-ray detectors demonstrated existence of subsequent powerful flares from the same source including one giant flare with $L\gtrsim 10^{44}$~erg~s$^{-1}$. A stable period of $\sim 8$~s was found and the object was localised in a SNR in the Large Magellanic cloud (LMC).
Now the McGill catalogue of magnetars\footnote{http://www.physics.mcgill.ca/~pulsar/magnetar/main.html} lists about 30 sources \cite{2014ApJS..212....6O}. They belong to two main subclasses: anomalous X-ray pulsars (AXPs) and soft gamma-ray repeaters (SGRs). A recent review with large bibliography  dedicated specifically to these types of sources can be found in \cite{2021ASSL..461...97E}.

Initially, division into AXPs and SGRs has been very clear. Anomalous X-ray pulsars were characterized by relatively stable X-ray emission with luminosity $\sim 10^{35}$~erg~s$^{-1}$ (i.e., substantially smaller than in most of accreting X-ray pulsars in binary systems); they had spin periods about a few seconds, which were always increasing; they did not have any optical or IR counterparts. Soft gamma-ray repeaters, in the first place, were characterised by intense bursts observed in hard X-rays and/or in soft gamma-ray range, AXPs did not show this type of activity. However, step by step it became clear that AXPs and SGRs share similar properties. Out of active periods SGRs often resemble AXPs. And in 2002 Gavriil et al. demonstrated that well-known AXPs can have bursting activity identical to that of SGRs \cite{2002Natur.419..142G}.   

Typical spin periods of magnetars are about few seconds. However, there are several important examples of outliers. A high-B young pulsar in a Crab-like plerion PSR J1846-0258 which started to demonstrate SGR-like flares \cite{2008Sci...319.1802G} has spin period 0.327~s. Oppositely, the source 1E 161348-5055 in the SNR RCW 103 (which I described above) has spin period $\sim 6.67$~hours. 

Situation with magnetic fields is also not so univocal. ``Classical'' magnetars have fields $\sim10^{14}$~--~$10^{15}$~G. However, there are several so-call ``low-field magnetars''. Up to now three objects  are reported (see a review in \cite{2013IJMPD..2230024T}). According to estimates based on the usual magneto-dipole equation, these sources has dipole field well below $10^{13}$~G. But phase-resolved spectroscopy demonstrated existence of proton cyclotron lines which indicates local surface fields $\sim10^{14}$~--~$10^{15}$~G \cite{2013Natur.500..312T, 2016MNRAS.456.4145R}. These might be small-scale non-dipolar components of the magnetic field. 

The best definition of a magnetar involves magnetic field dissipation. That is, a magnetar is not just a NS with large field, but such a compact object that magnetic energy release dominates in its luminosity at least for some period of time. The total energy budget can be roughly estimated as follows:

\begin{equation}
E_\mathrm{mag}\sim \frac43 \pi R_\mathrm{NS}^3 \left( \frac{B^2}{8\, \pi}\right) = 1.7 \times 10^{47}\, B_{15}^2\, \mathrm{erg}.
\end{equation}

 Naive estimates of magnetar ages based on the characteristic age $\tau_\mathrm{ch}\sim P/(2\dot P)$ are not valid as this simple equation is written for a constant field. However, association of some of the magnetars with SNRs and there position in the Galaxy \cite{2021MNRAS.508.1929C} robustly confirm their young ages $\sim 10^3$~--~$10^5$~yrs. 
 Young ages of magnetars indicate that an active period of magnetic energy dissipation does not last long. 


Magnetars might constitute a significant fraction of young NSs. 
Many studies indicate their fraction $\lesssim10$\%, see e.g. \cite{2010MNRAS.401.2675P} and references therein. However, at least one study suggests a much higher fraction of these objects: $\sim 40$\% \cite{2019MNRAS.487.1426B}.
The question of the magnetar fraction is closely related to the problem of formation of these NSs.
 

It is still unknown what defines if a newborn NSs is a magnetar. 
In one framework it is necessary to have a strongly magnetized progenitor \cite{2006MNRAS.367.1323F}. 
In another, the magnetic field is amplified by several orders of magnitude via a dynamo mechanism operating in a newborn NS \cite{1993ApJ...408..194T, 2006A&A...451.1049B}. 

In both scenarios it is quite probable that evolution in binaries can play a role. For example, observations of the magnetic star $\tau$~Sco suggest that its magnetic field was substantially increased due to coalescence of two main sequence stars in a  binary, and with magnetic flux conservation this star can become a magnetar in future \cite{2019Natur.574..211S}. On other hand, evolution in binaries can result in significant spin-up of the stellar core which later might make the dynamo mechanism efficient enough to produce a magnetar-scale magnetic field \cite{2016A&AT...29..183P}.

Rapid rotation which is necessary for production of large dipolar fields with the dynamo mechanism \cite{2020SciA....6.2732R} can be obtained in different ways. 
For example, a newborn NS can be spun-up due to fallback accretion \cite{2022A&A...668A..79B}.

Modeling of the magnetar evolution demonstrates \cite{2019LRCA....5....3P, 2021Univ....7..351I} that after some time $\sim10^4$~--~$10^5$~yrs the rate of magnetic energy dissipation decreases, all types of activity of a NS ceases. Thus, magnetars become sources of a different type. Most probably, one of their descendants are X-ray dim isolated NSs (XDINS), also known as the Magnificent seven (M7).


\subsection{Magnificent seven}

The M7 (or XDINSs) is group of near-by ($\lesssim$~a few hundred pc, see \cite{2007Ap&SS.308..171P}) young isolated NSs observed due to their thermal surface emission, see a review in \cite{2009ASSL..357..141T, 2020MNRAS.496.5052P}. The first member of this class of NSs was discovered in 1996 \cite{1996Natur.379..233W}.

The first period of the history of studies of the M7 is dominated by results from the ROSAT satellite, see
\cite{2001ASPC..234..225T} for an early brief review and discussion (early ideas included, e.g., the possibility that these NSs can be accreting sources with decayed magnetic field \cite{1997AstL...23..498K}).  
Since then many observations in different wavelengths were obtained (in addition to X-rays, many sources are detected as dim optical sources with magnitudes $\sim26$~--~28 and in near-UV, near-IR; in radio deep upper limits are obtained \cite{2009ApJ...702..692K, 2023arXiv230105509P}). 

Now, for all but one of these sources spin periods and their derivatives are measured, see e.g., Table 5 in \cite{2014A&A...563A..50P}.  The magneto-dipole formula provides an estimate of the magnetic field $\sim10^{13}$~--~$10^{14}$~G. 
The observed thermal emission can be either due to the residual heat, e.g. \cite{2006PhRvC..74b5803P}, or there is some contribution from  magnetic field decay \cite{2010MNRAS.401.2675P}.

Evolutionary, M7 might be descendants of magnetars \cite{2010MNRAS.401.2675P}. 
It was shown by population synthesis modeling that the population of the M7 originated mainly from the Gould Belt -- the local ($\sim500 $~pc) starforming structure \cite{2005Ap&SS.299..117P}. As these NSs are relatively weak ($L\sim 10^{31}$~--~$10^{32}$~erg~s$^{-1}$) and soft ($kT\lesssim 100$~eV), it is difficult to detect such sources at large distances, mostly due to the interstellar absorption.

Despite intensive searches (e.g., \cite{2011AJ....141..176A} and references therein) very few NSs similar to the M7 were found and none of them ideally resembles the original seven sources. 
The first one was Calvera -- a soft X-ray source high above the Galactic plane \cite{2009ApJ...705..391S}. But later it was shown that this NS has a short spin period (0.06 s) and also in some other respects is different from the M7 sources \cite{2021ApJ...922..253M}.
The next one is 2XMM J104608.7-594306 \cite{2009A&A...498..233P}. Again, the spin period (18.6 msec) does not fit the M7 family \cite{2015A&A...583A.117P}. Finally, the latest discovery is the source 4XMM J022141.5-735632  \cite{2022A&A...666A.148P}. For this object the spin period is not reported, yet.

There have been hopes that the eROSITA telescope can find much more M7-like sources \cite{2018IAUS..337..112P, 2021ARep...65..615K}, but as it has been switched off just after two years of the survey program these hopes more or less disappeared.

\section{Standard evolution and its problems}

 It is convenient to discuss evolution of young NSs in terms of $P$, $\dot P$, and $B$; and to illustrate it with the $P$~--~$\dot P$ diagram. In the simplest and the most standard way the evolution is described by Eq.~(1) for $\mu =$const.  Then NSs evolve in the $P$~--~$\dot P$ diagram along strait tracks, Fig.~1. 
 
 Absence of significant field decay in normal PSRs was found in many papers, e.g. \cite{2006ApJ...643..332F} (see, however, the next section). Also, in PSRs we do not see any evidence for additional release of magnetic energy (situation with magnetars is drastically different, of course).

 Typically, in this standard approach it is assumed that the initial spin periods are very short. 
Sometimes authors can assume that the initial period is close to the limiting rotation (e.g., 1 ms).
Sometimes, these initial spin periods are assumed to be close to the initial period of the Crab pulsar. Such assumptions were very popular, for example, in early models of binary population synthesis, e.g. \cite{1996smbs.book.....L}. Due to gradual progress in our understanding of initial parameters of NSs such simplified approaches were replaced by more advanced ones.

 If $P_0\ll P$ and the field is constant then the real age of a pulsar is close to the characteristic age $\tau_\mathrm{ch}$. Population synthesis studies and analysis of young NSs in SNRs with known ages indicate that typical initial periods of the majority of PSRs are of the order of 0.1~s \cite{2006ApJ...643..332F, 2012Ap&SS.341..457P, 2022MNRAS.514.4606I}. Thus, for many standard PSRs with observed $P\sim 1$~s the assumption of small initial period can be acceptable. Still, in many cases it does not work well and results e.g., in a significant discrepancy between the real age and $\tau_\mathrm{ch}$.

 The simplest model of magneto-rotational evolution nearly excludes links between different subpopulations: a CCO cannot become an M7-like object, and a magnetar cannot appear later in its life as a standard radio pulsar. This feature leads to an interesting controversy. The sum of birthrates of different subpopulations is larger than the rate of CCSN \cite{2008MNRAS.391.2009K} (note, in that paper the authors do not include CCOs in their calculations, with this subpopulation the problem is even more severe).  
 Also, in the simplest model it is difficult to explain lack of magnetars with periods larger than $\sim 10$~s, e.g. \cite{2000ApJ...529L..29C}.  As well as the absence of descendants of CCOs which might be visible at ages $\lesssim 10^6$~yrs when a SNR is already dissolved \cite{2012ASPC..466..191P}. 

 Thus, it necessary to consider more complicated evolutionary paths.

\section{Double nature and non-standard evolution}



In this section I discuss two possible features of NSs evolution: magnetic field decay and fallback.
Simplified tracks are shown in Fig.~2. 
Also I present several examples of sources which absolutely do not fit the simplified NS evolution but require either field decay, or fallback, or both.

\begin{figure}[H]
\includegraphics[width=10.5 cm]{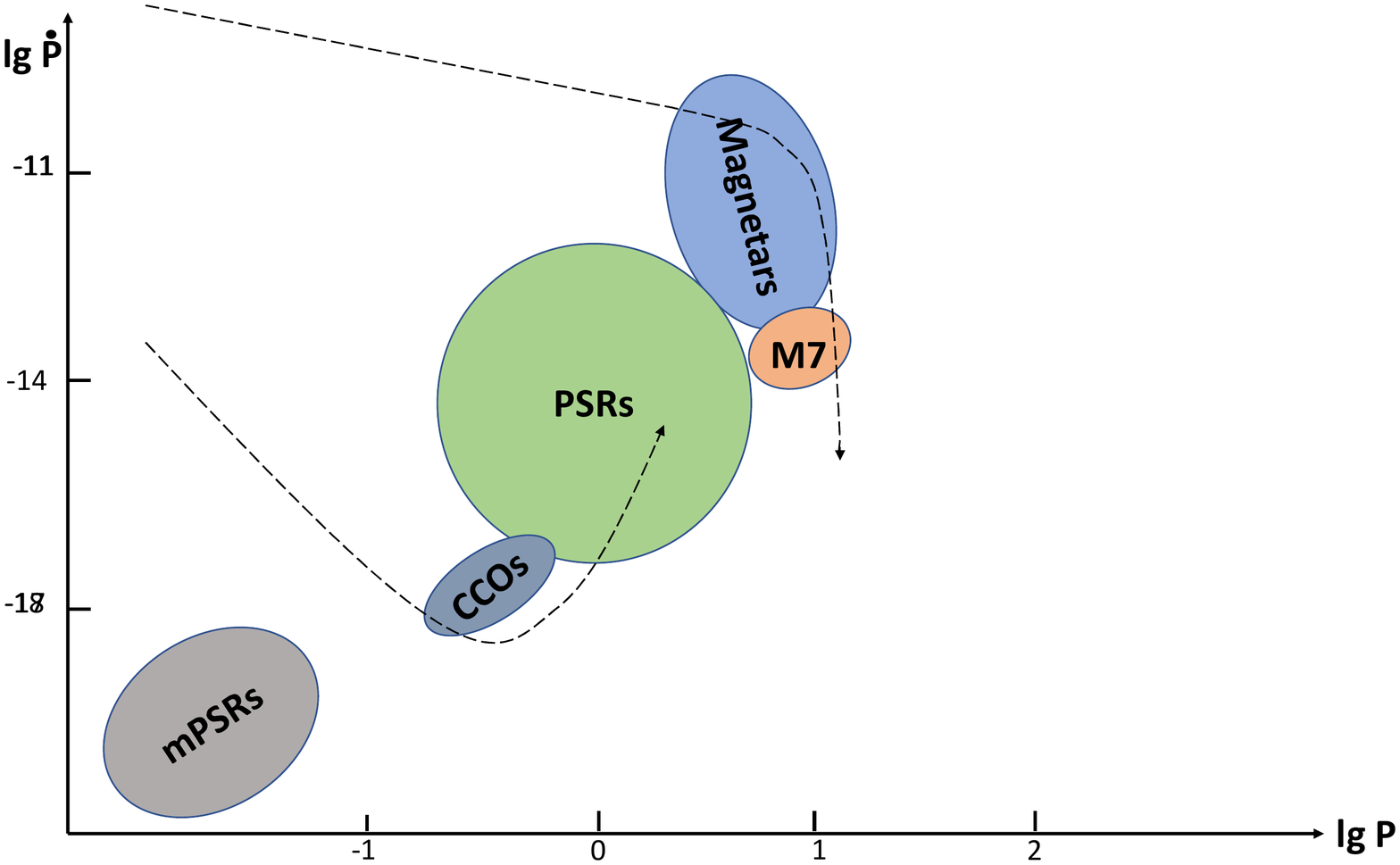}
\caption{The same as Fig,~1, but with evolutionary tracks corresponding to field decay in magnetars and re-emerging fields in CCOs. \label{fig12}}
\end{figure}

\subsection{Magnetic field decay}

It is quite natural to expect that magnetic fields of NSs might decay in time. 
In addition to obvious general physics arguments that electric currents which are responsible for the magnetic field might decay due to finite conductivity in the crust (and maybe due to other processes, like transport of magnetic flux tubes from the core to the crust, e.g. \cite{1998ApJ...492..267R}) there are observational arguments. The existence of low fields in old NSs was demonstrated e.g., by discovery of mPSRs \cite{1982Natur.300..615B}. 

 Theory of magnetic field decay in NSs and related observational data were reviewed many times, see e.g. 
 \cite{2019LRCA....5....3P, 2021Univ....7..351I} and references therein. Here I just briefly remind several basic features. 

Magnetic field can exist in a solid NS crust or/and in a liquid (and, most probably, superconducting) core. In the first case the magnetic field is produced by electric currents. In the second, the field is confined in magnetic flux tubes, as the core is expected to be type-II superconductor, see e.g. \cite{1969Natur.224..673B}.  Thus, physics of the field evolution is very different in these two cases. 

Physics of the core is much less understood. Partly because of that often only field evolution in the crust is considered. Basics of crustal field evolution are perfectly described in \cite{2004ApJ...609..999C}.
Two main time scales can be defined: the Ohmic ($\tau_\mathrm{Ohm}$) and the Hall one ($\tau_\mathrm{Hall}$).  

The Ohmic scale can be written as:
\begin{equation}
\tau_\mathrm{Ohm} = \frac{4\pi \sigma L^2}{c^2} .
\end{equation}
Here $\sigma$ is conductivity and $L$ is a length scale in the crust. 

The Ohmic time scale depends on  temperature of the crust as electrons can scatter off phonons, and on  crustal composition (impurities). For usual field configurations, $L$ is sufficiently large ($\sim$ a few hundred meters in deeper layers, comparable to the crust thickness, see \cite{2004ApJ...609..999C}) to make the time scale relatively long. In young hot NSs $\tau_\mathrm{Ohm}$ can be about $10^5$ yrs, 
see e.g. Fig.~4 in \cite{2015AN....336..831I}. And in older cold NSs this time scale is long $\sim10^9$~yrs. 

Rapid release of magnetic energy can proceed via the Hall cascade \cite{1992ApJ...395..250G}. This is a non-dissipative process.
But it reconfigures the field such that $L$ decreases, so that now the field can  decay faster. 
The time scale of this process is:
\begin{equation}
\tau_\mathrm{Hall} = \frac{4\pi e n_eL^2}{cB(t)} ,
\end{equation}
where $e$ is the elementary charge and $n_e$ is concentration of electrons.
For fields $\sim 10^{15}$~G it can be as small as $\sim 100$~yrs, e.g. \cite{2008ApJ...673L.167A}. 
However, the value of $ \tau_\mathrm{Hall}$ is not well-known, and it can be two orders of magnitude large for the same field $10^{15}$~G.
It is widely accepted that magnetars activity is related to the Hall cascade in NS crust.

Evolution of magnetic fields in a core is described in a more complicated way (see a brief review is Sec.3.2 in \cite{2021Univ....7..351I}). 
Recently, Gusakov, Kantor, and Ofengeim developed a new approach to calculate field behaviour in superconducting cores
 \cite{2017PhRvD..96j3012G, 2020MNRAS.499.4561G}. 
In particular, their results suggest that in vicinity if the crust the time scale can be as short as $\sim 100$~yrs \cite{2018PhRvD..98d3007O}. This is intriguing as it potentially gives an opportunity to explain magnetar activity in the framework of the core field evolution.

From the observational point of view there are many arguments in favour of decaying fields in young NSs of different types. In the first place, magnetar activity provides evidence for the field decay, as obviously the magnetic energy is released in bursts and it is responsible for the crust heating. Active lifetime of magnetars might be short as it comes out from independent ages estimates of these sources (SNR and kinematic ages, associations with clusters of young stars, etc.). 
However, bursts can be produced also by older NSs, but more seldom
\cite{2011ApJ...727L..51P}. 

Thermal properties of NSs also provide arguments in favour of the field decay. E.g., Pons et al.
\cite{2007PhRvL..98g1101P} demonstrated that typically a high-B NS cannot have low surface temperature due to additional heating related to the magnetic energy release in the crust. 

Analysis of properties of high mass X-ray binaries (HMXBs) showed that distribution of magnetic fields of NSs in these systems is compatible with models of crustal field evolution
\cite{2012NewA...17..594C}.

Finally, even for normal PSRs some modeling favoured decaying field along their evolution, see e.g., \cite{2002ApJ...565..482G} and references therein. A different conclusion was made in
\cite{2014MNRAS.444.1066I}. These authors constructed a modified model of so-called ``pulsar current'' \cite{1981MNRAS.194..137P, 1981JApA....2..315V} and concluded that a significant fraction of normal pulsars experiences an episode of field decay with a time scale $\sim 4\times 10^5$~yrs at ages $\lesssim 10^6$~yrs. Later this decay might be terminated. This points to the Ohmic decay due to electron scattering off phonons as this type of decay disappears when a NS becomes sufficiently cold. This happens at ages $\lesssim 10^6$~yrs even for low-mass objects. In the same framework anomalous braking indices of PSRs can be explained, too \cite{2020MNRAS.499.2826I}.

To summarize, magnetic field decay in NSs is now a standard ingredient of modeling their evolution. 
Different modes of decay with different time scales are possible. So, presently the situation is far from being clear. That is why observations of peculiar objects are important, and I discuss some of them in the following subsection.
 
\subsection{Werewolves and secret agents}

 Till the beginning of this century it was possible to attribute each young NS to some  well-defined category (PSR, AXP, SGR, CCO, M7, etc.). In 2002 the discovery of SGR-like bursts from an AXP has been announced \cite{2002Natur.419..142G}. That was the first, but not very prominent example of ``double nature''. Not so prominent because for some time is had been already suspected that AXPs and SGRs form the same family of objects -- magnetars. May be, SGRs are slightly younger -- and so, more active. 
But later on more pronounced examples of transition from one subpopulation to another were found. 
Here I give some examples.

PSR 1846-0258 is observed only as an X-ray source -- the radio beam is not pointing towards the Earth.
The NS has $B\sim 5\times 10^{13}$~G and the largest rotational energy losses $\dot E_\mathrm{rot}$
among PSRs. A plerion and a SNR are observed around the source. The characteristic age is $\sim 884$~yrs. The spin period is $\sim 0.33$~s. So, it looked like a Crab-like PSR with an order of magnitude large field and an order of magnitude longer period. But in 2008 a magnetar-like activity was reported from this object
\cite{2008arXiv0802.1242S, 2008Sci...319.1802G}. 
X-ray luminosity of the PSR significantly increased and it started to produce SGR-like bursts. 
This came out to be the first example, when a radio pulsar became a magnetar. 

Another example of pulsar $\rightarrow$ magnetar transition is PSR 1622-4950.
It has $P=4.3$~s and a large period derivative corresponding to $B\sim 3 \times 10^{14}$~G. 
Since its discovery it has been suspected
\cite{2010ApJ...721L..33L} that the source can be a magnetar in a quiescent state.
Indeed, in 2017 the source re-activated
\cite{2018ApJ...856..180C}. Its X-ray luminosity significantly increased, however, no bursts were detected. 

Above I already described a very peculiar source in the SNR RCW 103 which initially looked like an atypical CCO, but then appeared to be an active magnetar.
Its activity continued in 2016 with an outburst \cite{2019A&A...626A..19E} which then decayed following the general scenario of crustal magnetic energy release.

PSR J1852+0040 in Kes 79 also was presented above. It is a candidate to ``hidden'' magnetars, i.e. a magnetar covered during a fallback episode such that only its crustal (but not magnetospheric) activity can be visible. This NS has a spin period $\sim 0.1$~s \cite{2010ApJ...709..436H}.  If we assume that the magnetar scale field was buried by fallback, then the initial episode has been very rapid so that the NS had no time to increase the spin period due to interaction with the fallback flow (as it, most probably, also happened in RCW 103). Then, the spin period of PSR J1852+0040 could be ``frozen''. So, the present day value can be similar to the initial spin. Then, it means that magnetar formation does not necessary require rotation with a rate much smaller than 0.1~s \cite{2013PASA...30...45P}.
It is tempting to say that a compact remnant of the SN 1987A  also can be a ``hidden'' magnetar \cite{2015PASA...32...18P} as its progenitor was a product of a coalescence of two massive stars in a binary system \cite{2007Sci...315.1103M}. This coalescence could enhance spin rate and magnetic field of the stellar core of the progenitor. Discovery of more ``frozen'' magnetars can shed light on their initial rotation rate and so -- on the mechanism of magnetar formation.  

After a fallback episode the field is expected to diffuse out on a time scale $\sim 10^3$~--~$10^5$~yrs
\cite{2011MNRAS.414.2567H}. 
While the field is diffusing out, its external structure can be changed which can prevent appearance of radio pulsar emission \cite{2016MNRAS.462.3689I}. 
Still, there is a possibility to observe a PSR on the stage of field re-emergence. Then, it might have a very non-standard track in the $P$~--~$\dot P$ diagram. And such objects are known! 
The most famous example is PSR 1734-3333 \cite{2011ApJ...741L..13E}. 
This is a standard PSR, but its period derivative is rapidly increasing. The rate corresponds to the braking index $\sim 1$, $\dot P \propto P^{2-n}$ (the standard Eq.~1 corresponds to $n=3$). With $P=1.17$~s and a large period derivative $\dot P=2.28\times 10^{-12}$~s/s it can after some time -- $\sim 20$~--~30~kyr, -- enter the region of magnetars, see \cite{2011ApJ...741L..13E}.

 Finally, it is necessary to say few words about low-field magnetars which were also mentioned above. 
 As proposed in \cite{2014ApJ...781L..17R} these sources can form a significant subpopulation of relatively old magnetars. These sources might demonstrate activity just very seldom and have low quiescent luminosities. 
 Their existence points to the possibility of quasistationary configuration of magnetic field with relatively low dipolar component. In the following subsection I discuss one of recently proposed stable field configurations -- the Hall attractor.

\subsection{Fallback and Hall attractor}

Examples of peculiar sources described above suggest that NS field evolution can follow non-standard routes. In this subsection we consider two important features of such evolution: fallback and Hall attractor. 

The idea that some fraction of matter ejected after a bounce in a SN explosion can later fallback onto the NS was proposed in early 1970s (see a brief historical review in the introductory section of \cite{1989ApJ...346..847C}).
In 1995 Muslimov and Page \cite{1995ApJ...440L..77M} suggested that fallback can significantly influence external magnetic field of NSs and delay switch-on of radio pulsar emission mechanism. 
 
The scenario with magnetic field submergence due to fallback became popular when it was applied to CCOs by Ho
\cite{2011MNRAS.414.2567H} and then by Vigano and Pons
\cite{2012MNRAS.425.2487V}. These authors demonstrated that for a realistic fallback amount  ($\Delta M \sim 10^{-6}-10^{-4}\, M_\odot$) magnetic field can be significantly submerged and diffuses out on the time scale $\sim 10^3$~--~$10^4$~yrs. This perfectly fits properties of CCOs and explains why ``evolved CCOs'' are not observed as purely thermal emitters with $P\lesssim 1$~s, e.g. \cite{2012ASPC..466..191P}.

Bernal et al. presented 2D and 3D simulations of magnetic field submergence due to fallback
\cite{2013ApJ...770..106B}. They modeled dynamics of interaction between falling matter and magnetic field on the scale $\lesssim 100$~msec for different fallback rates.  The authors show that for  rates $\dot M\gtrsim 10 \, M_\odot$~yr$^{-1} $ the total submergence happens, for lower rates the field is submerged just partially.  Such rates are realistic at early stages of fallback, so in some fraction of young NSs the external magnetic field can be smaller than the crustal field by several orders of magnitude.

Fallback can be prevented by activity of the central source. This possibility has been neglected in e.g. \cite{2013ApJ...770..106B}, but later on it was  studied by the Japanese group 
\cite{2018PASJ...70..107S},
\cite{2021ApJ...917...71Z}. These authors attribute diversity of young NSs mainly to different amount of fallback. In particular, in \cite{2018PASJ...70..107S} they define criteria according to which in a simplified 1D model for a given fallback rate a NS becomes a CCO, a PSR, or a magnetar depending on its spin and magnetic field. In \cite{2021ApJ...917...71Z} the same situation was studied in more details with a numerical approach, but again in the 1D approximation and without accounting for instabilities. 

Advanced fallback calculations are performed in a framework of SN explosion modeling
\cite{2022ApJ...926....9J}. 
A specific feature of this modeling is motion of the NS relative to the ejecta. It is shown, that this results in spin-up of a newborn NS due to fallback and in spin-velocity alignment. In this framework a NS spin in mostly determined by the fallback (and not by the spin rate of the progenitor). 

Influence of fallback on the spin of a newborn NS is studied also in
\cite{2022A&A...668A..79B}. In this scenario a NS formed from a slowly rotating progenitor star can be spun-up so significantly that conditions necessary for a magnetar formation are fulfilled. 
Thus, in this model fallback also produces rapidly rotating compact objects. An opposite situation is also possible in other scenarios.

Interaction between a fallback disc and magnetic field of a magnetar can result in significant spin-down of the NS. This possibility was recently analysed in
\cite{2022ApJ...934..184R}. For a wide range of realistic fallback rates the disc can penetrate within the light cylinder. Thus, the NS enters the propeller stage of magneto-rotational evolution. At this stage a compact object can spin-down rapidly. Periods $\sim10^2$~--~$10^4$~s can be easily reached even within a lifetime of a SNR ($\lesssim 10^5$~yrs). 
This scenario is applicable for recently discovered long-period pulsars, discussed in the following section. 

Long spin periods also can be reached if magnetic field remains large for a long time. 
This is possible if the Hall cascade in a magnetar crust is terminated or at least significantly slowed down. Such situation has been found numerically and the stage was named ``the Hall attractor'' \cite{Gourgouliatos2013MNRAS, 2014PhRvL.112q1101G}. Later, it was confirmed in \cite{2015PhRvL.114s1101W, Bransgrove2018MNRAS}.

In the original paper \cite{2014PhRvL.112q1101G} the authors obtained that the attractor is reached in $\lesssim 1$~Myr for initial fields $\sim 10^{14}$~G. For larger fields it is reached faster. At the attractor stage the dipolar field is about $\exp(-3) \times B_0$, where $B_0$ is the initial field. Details significantly depend on the model, in particular -- on the initial conditions (see a review of magnetic field evolution in NSs in \cite{2021Univ....7..351I}).  The bottom line is the following: rapid initial Hall evolution of large magnetic fields can be significantly slowed, this potentially allows existence of NSs with relatively large fields at ages at least $\sim 1$~Myr (which is also important for explanation of magnetar candidates in accreting binary systems \cite{2018MNRAS.473.3204I}). In this case such objects can reach relatively large spin periods just due to standard losses. This can help to explain some sources discussed in the next section.

\section{New puzzle -- new tracks}

 Recent discoveries of long spin period pulsars demand new non-trivial evolutionary tracks in comparison with those shown in Fig.~2. 

MeerKAT observations allowed Caleb et al. to discover a radio pulsar PSR J0901-4046 with a record-long spin period 76 s  \cite{2022NatAs...6..828C}.
With $\dot P=2.25\times 10^{-13}$~s~s$^{-1}$ the source has the characteristic age 5.3~Myr. The magneto-dipole field estimate provides the value $1.3 \times 10^{14}$~G. In the standard scenario of magneto-rotational evolution such combination of parameters is impossible due to short initial spin periods and rapid decay of large magnetic fields. 

GLEAM-X J162759.5-523504.3 is even more exotic with the period of pulsations of its radio emission $\sim18$ minutes
 \cite{2022Natur.601..526H}.
This object was discovered wiht a help of the Murchison Widefield Array. The period derivative is not measured, yet. This prevents robust determination of the source nature. Still, most probably it is a NS not a WD, see discussion and references in \cite{2023MNRAS.520.1872B}.
Moreover, it might be a magnetar as an upper limit on $\dot P \lesssim 10^{-9}$~s~s$^{-1}$ provides that the observed luminosity is larger than the rotational energy losses. Thus, an additional source of energy is necessary and it can be magnetic energy of the magnetar.

If $\dot P $ of GLEAM-X J162759.5-523504.3 is close to the upper limit then the dipolar field is $\sim 3\times 10^{16}$~G.  Such values have been never observed. The characteristic age for such field is $\tau_\mathrm{ch} \sim   10^4$~yrs which is significantly larger that the expected time scale of Hall cascade for so huge fields.  If $\dot P \sim 10^{12}$~s~s$^{-1}$ then the field is $\sim 10^{15}$~G and $\tau_\mathrm{ch} \gtrsim 10^7$~yrs. Again, such combination is not a part of the standard scenario of NS evolution. 

\begin{figure}[H]
\includegraphics[width=10.5 cm]{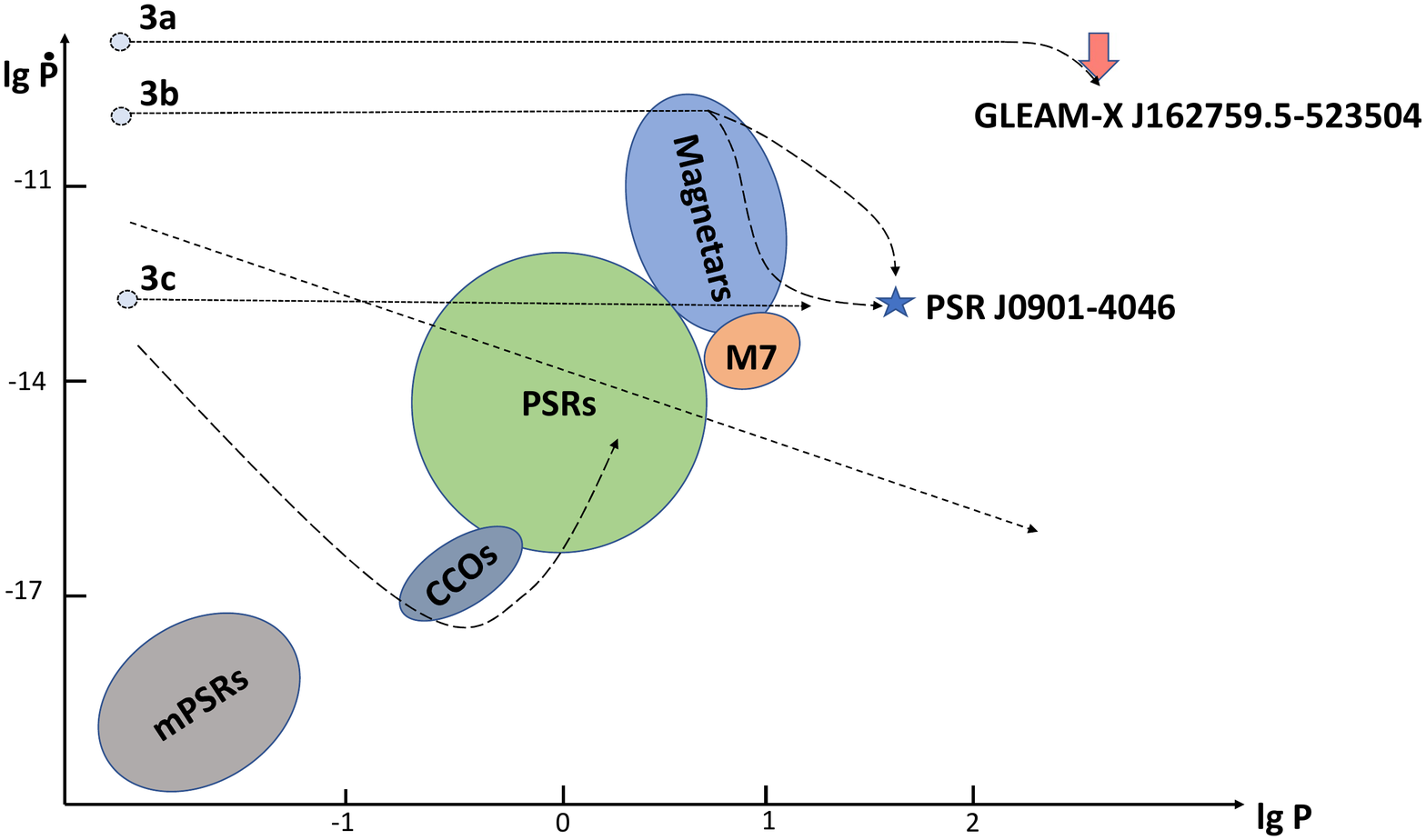}
\caption{The same as Fig,~2, but with recently discovered long period pulsars (the star symbol -- PSR J0901-4046, the arrow corresponds to the spin period and the upper limit on the period derivative of GLEAM-X 162759.5-523504) and tracks which can illustrate their evolution, see text for details.  
\label{fig3}}
\end{figure}

Possible solutions are related to physical processes discussed in the previous section. 
Either real ages of both objects are much smaller than their characteristic ages due to large initial spin periods, or the dipolar magnetic field in both cases could survive for a very long time, much longer than the initial time scale of the Hall cascade. In Fig.~3. we show the tracks with large initial spin periods. The initial short-dashed part of the tracks pointing towards PSR J0901-4046 and 
GLEAM-X J162759.5-523504.3  corresponds to a rapid spin down from short period just after the core collapse to longer periods, e.g. due to interaction with the fallback disc (of course, this spin-down does not proceed with constant period derivative, so do not take this part of the tracks literally).  
For  PSR J0901-4046 two variants of the further evolution are shown: standard field decay and stalled decay.

Note, that above we discussed only scenarios involving single stars. Evolution in a binary system can open an additional channel of producing long spin periods of NSs. If a NS in a HMXB system rapidly starts to accrete \cite{2022MNRAS.511.4447K}, or at least reaches the propeller stage, then its period can be rapidly increased up to hundreds or thousand of seconds in case of large magnetic fields and accretion from a stellar wind, see a catalogue of HMXBs in \cite{2023A&A...671A.149F}. If this happens close to the moment of explosion of the secondary component then we can expect a ``birth'' of an isolated NS with a large spin period. 

 NSs which can rapidly reach long spin periods (and which, probably, save large value of their dipolar magnetic fields for a long time) can be of special interest for a long-term evolution of NSs. I discuss it in the following section. 

\section{Towards accretion from the ISM}

Already more than 50 years ago it was suggested that isolated NSs sooner or later can start to accrete gas from the interstellar medium (ISM) \cite{1970ApL.....6..179O, 1971SvA....14..662S}. 
More than 30 years ago it has been proposed that e.g., ROSAT could detect thousands of accreting isolated NSs (AINSs) \cite{1991A&A...241..107T}, but none were found. This was explained in \cite{2000ApJ...530..896P} as an evolutionary effect: most of isolated NSs under the standard assumptions cannot reach the stage of accretion during lifetime of the Galaxy. In addition, the rate of accretion onto surface of a NS can be much lower than the standard Bondi value $\dot M \propto \eta \frac{(GM)^2}{v^3} \rho \sim 10^{11} \, \left(\frac{10\, \mathrm{km\, s^{-1}}}{v}\right)^3 \left(\frac{\rho}{10^{-24}\, \mathrm{g \, cm^{-3}}}\right)\, \mathrm{g \, s^{-1}}$ due to magnetic inhibition \cite{2003ApJ...593..472T}. In the formula $v$ is the NS velocity relative to the ISM and $\rho$ is the ISM density, the coefficient $\eta\sim 10$ depends on details of accretion flow around the NS. 

Velocity distribution is an important ingredient of isolated NS evolution modeling as interaction of a compact object with the interstellar medium strongly depends on this parameter. Also, the initial velocity distribution determines spatial distribution of NSs in the Galaxy, see e.g. \cite{2022MNRAS.516.4971S}.
Already early observations  demonstrated that NSs can have spatial velocities significantly large than their progenitors \cite{1970SvA....13..562S}. It is assumed that NSs obtain an additional velocity at birth (so-called ``kick''). The origin of kick, shape of the velocity distribution, and possible correlations of the kick velocity with other parameters are not completely understood, yet. During last $\lesssim 50$~yrs many attempt were made to derive the kick velocity distribution from observations or to obtain it from theoretical considerations, e.g. SN explosion models (see a brief review and references to early studies in the introductory part of \cite{2022MNRAS.517.3938C}). It is quite popular to use bimodal velocity distributions as they fit better various data on radio pulsars and X-ray binaries (especially those with a Be-star donor).
Recently, in \cite{2021MNRAS.508.3345I} the authors presented a new analysis where they investigated properties of radio pulsars and HMXBs. Their best fit is a bimodal distribution with $\sigma_1\sim$~30--70~km~s$^{-1}$ and $\sigma_2= 336$~km~s$^{-1}$, where 10-30\% of NSs come from the low velocity component.
If isolated, such NSs can become potentially observable accreting sources within the Galactic life time.

NSs with larger magnetic fields can start to accrete faster. This was studied in detail in
 \cite{2010MNRAS.407.1090B}. 
The problem of low accretion luminosity can be solved in the settling accretion scenario \cite{2012MNRAS.420..216S}. In this framework an accreting isolated NS can be observed as a relatively bright (e.g., for eROSITA) transient source \cite{2015MNRAS.447.2817P}. But still, the number of isolated accretors is not expected to be very high which makes their searches problematic.

Discovery of isolated accreting NSs is very much welcomed as it can open a unique possibility to study old isolated NSs and to learn a lot about their properties and evolution. eROSITA could be a perfect instrument to reach this goal \cite{2021ARep...65..615K}. On other hand, it is important to provide better estimates of the number of accreting isolated NSs and their properties in order to simplify identification of these objects. 

\begin{figure}[H]
\includegraphics[width=10.5 cm]{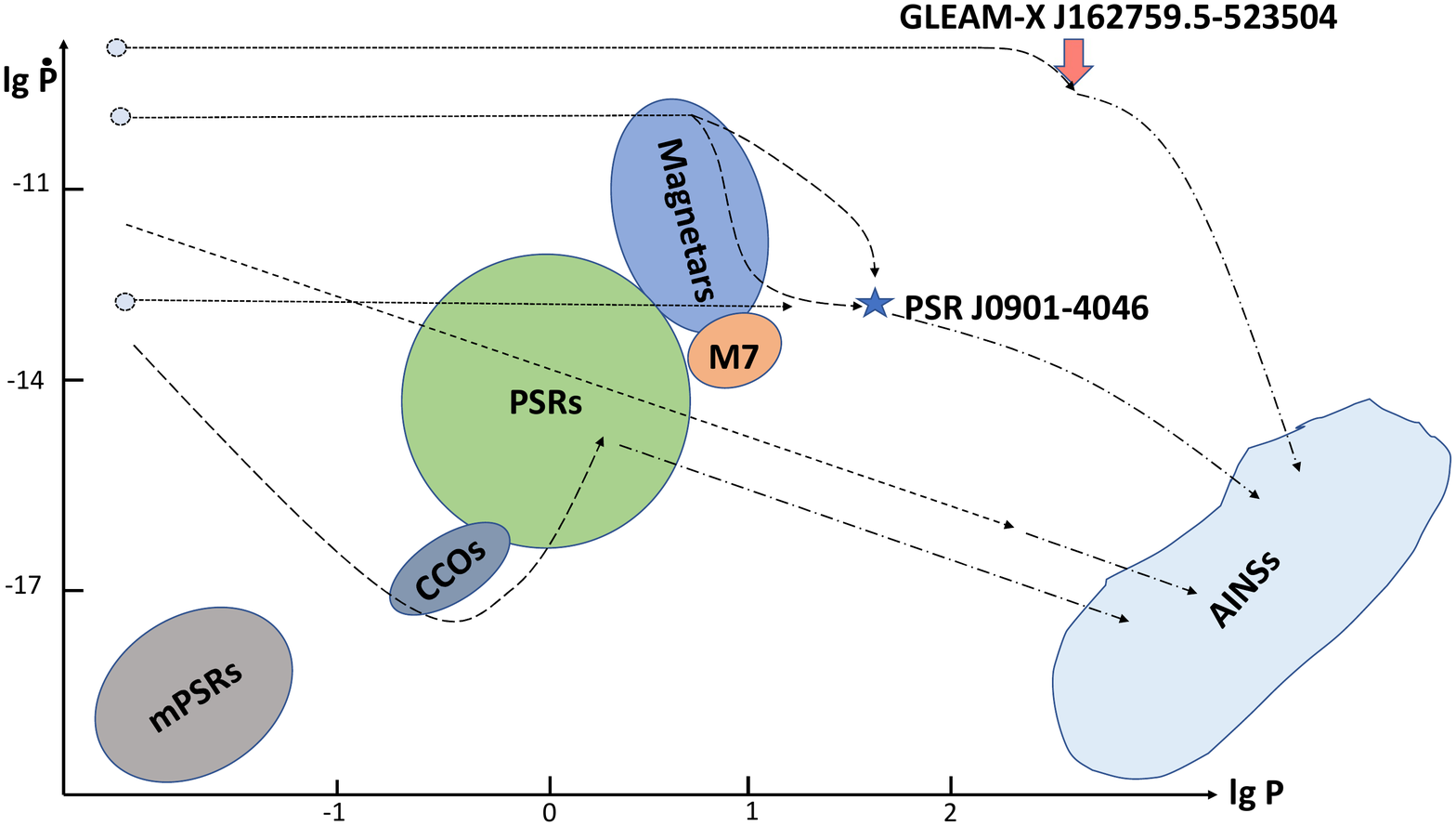}
\caption{The same as Fig,~3, but with addition of the region of AINSs and illustration of corresponding evolutionary tracks. See text for details. Position of AINSs is added out of scale. \label{fig4}}
\end{figure}

In the standard approach \cite{2000ApJ...530..896P} the main obstacle on the way to accretion is related to relatively slow spin-down of an isolated NS with a standard magnetic field $\sim 10^{12}$~G. 
Recently discovered young very long period NSs are good candidates to reach the stage of accretion in a relatively short time. Thus, estimation of the number of such objects is of great interest for long term NS evolution.

In general, all NSs with long spin periods, large long lived magnetic fields, and/or low spatial velocities (e.g., those born in e$^-$-capture SN) have good chances to reach the stage of accretion from the ISM. 

\section{Magnetars and FRBs}

Fast radio bursts (FRBs) are millisecond-scale radio transients discovered in 2007 \cite{2007Sci...318..777L}, see a recent comprehensive review in \cite{2022arXiv221203972Z}.
A possible link to NSs, in particular -- to magnetars, -- was proposed already in 2007 \cite{2007arXiv0710.2006P}. In 2020 it was confirmed by detection of simultaneous radio and high-energy flares from the Galactic magnetar
SGR 1935+2154~\cite{2020Natur.587...54C, 2020Natur.587...59B, 2021NatAs...5..378L, 2020ApJ...898L..29M, 2021NatAs...5..372R, 2021NatAs...5..401T}.

The number of known sources of FRBs is rapidly growing and now it is about $\sim10^3$. About 50 of the known sources demonstrate repeating activity \cite{2023arXiv230108762T}, four of the repeaters show very high rate of events producing up to several hundred bursts per hour \cite{2022arXiv221205242H}. 
In near future FRBs might become the most numerous known sources related to NSs, what is also important -- they are extragalactic up to $z\gtrsim 1$. Thus, they will be one of the main sources of information about the universal population of NSs \cite{2023Parti...6..451P}.

NSs producing FRBs can have different peculiar properties and origin. 
In the first place, it is expected that FRB sources are extreme magnetars with large fields producing hyperflares with total energy release $\sim10^{44}$~erg and peak luminosities $\sim 10^{47}$~erg~s$^{-1}$, which correspond to a millisecond radio burst with $L\sim 10^{43}$~erg~s$^{-1}$ with a ratio $\frac{L_\mathrm{radio}}{L_\mathrm{total}}\sim 10^{-4}$ (see e.g. \cite{2021NatAs...5..372R}).

Four of the most active FRB sources demonstrate such a huge rate of flares (hundreds per hour) that from the energetic point of view such behavior cannot last longer than few years, as the whole magnetic energy $\gtrsim 10^{47}\left(\frac{B}{10^{15} \, \mathrm{G}}\right)^2$~erg, see Eq.~(2), would be emitted in this period \cite{2023arXiv230414665Z}. Such intense outbursts are not observed among Galactic magnetars. 

Two of the repeating sources of FRBs demonstrate periodicity on the scale $\sim 16$ \cite{2020Natur.582..351C} and $\sim 160$
\cite{2020MNRAS.495.3551R} days. The origin of this periodicity is unknown. Among the proposed hypotheses there are the following: binarity \cite{2020ApJ...893L..39L, 2022MNRAS.515.4217B}, NS precession \cite{2020ApJ...892L..15Z, 2020ApJ...895L..30L}, and extra-long spin periods \cite{2020MNRAS.496.3390B}.
All of these opportunities are very intriguing as we do not know robust examples of active magnetars in binary system (see a review in \cite{2023IAUS..363...61P}), we have just a few unconfirmed candidates for precessing magnetars (see e.g., \cite{2021MNRAS.502.2266M} and references therein), and we do not know any examples of so long spin periods of NSs. 

Emission mechanism of FRBs is not figured out, yet. Presently, two main frameworks are discussed: magnetospheric emission and external relativistic shocks,  see reviews in \cite{2020Natur.587...45Z, 2022arXiv221014268P}. Advanced theoretical scenarios are proposed in large number for both families of models. Growing variety of observational data (including polarization measurements, burst structure, spectra and their evolution during bursts) on the one hand poses many questions, and on the other hand -- provides lots of opportunities to test model predictions.
Probably, observations of simultaneous radio and X/$\gamma$-ray flares from Galactic magnetars will help to select the correct approach. Understanding of the origin of FRB radiation might shed light on important properties related to NS emission properties, in general.

The Galactic population of magnetars is consistent with an assumption that all these sources originated from core-collapse SN. Indeed, these NSs demonstrate clear correlation with young stellar populations and sometimes are situated inside standard SNRs, see e.g. \cite{2015RPPh...78k6901T} for a review.
However, a magnetar (or a NS, in general) can be formed via several other channels. Mostly, they are related to coalescence of compact objects: NSs or/and WDs. 

FRB sources are identified in different types of host galaxies in various environment \cite{2023arXiv230205465G}, including a source in a globular cluster \cite{2022Natur.602..585K}. 
Localisation of FRBs at sites of very low star formation points towards alternative evolutionary channels related to old stellar populations. Coalescence NS-NS, NS-WD, WD-WD altogether can produce NSs with a rate at most $\sim 10^{-4}$~yrs$^{-1}$ per a Milky way-like galaxy (see references in e.g., \cite{2023Parti...6..451P}). Thus, the probability to find at least one active magnetar with such origins in our Galaxy is not high. Observations of FRBs allow us to study these sources, even in different epochs of cosmic history. Moreover, in near future new sensitive low-frequency radio telescopes might allow to observe FRBs from objects originated from Pop III stars!  

Understanding properties of sources of FRBs can bring us new surprises about NS physics and  observational appearances.

\section{Conclusions}

 The field of NS astrophysics actively develops, in the first place thanks to discoveries of new peculiar sources (like long spin period pulsars) and types of sources (like FRBs). Phenomenology of NSs becomes richer and richer and this requires more advanced theoretical approaches. We see more and more evolutionary links between different beasts in the zoo of NSs. Understanding of this diverse population of sources is a fascinating task and we keep going on. 


\funding{SP acknowledges support from the Simons Foundation which made possible the visit to the ICTP.}

\acknowledgments{I am grateful to the Organizers of the 2nd International Electronic Conference on Universe (ECU 2023), and personally to  Nicholas Chamel, for the invitation to present a talk about different types of isolated NSs which became the basis for the present review.}

\conflictsofinterest{The author declare no conflict of interest.} 



\abbreviations{The following abbreviations are used in this manuscript:\\

\noindent 
\begin{tabular}{@{}ll}
AINS & Accreting isolated neutron star \\
AXP & Anomalous X-ray pulsar\\
BH & Black hole\\
CCO & Central compact object\\
FRB & Fast radio burst\\
HMXB & High-mass X-ray binary \\
ISM & Interstellar medium \\
LMXB & Low-mass X-ray binary \\
M7 & Magnificent seven\\
mPSR & Millisecond radio pulsar \\
NS & Neutron star\\
PSR & Radio pulsar\\
SGR & Soft gamma-ray repeater\\
sGRB & Short gamma-ray burst \\
SN & Supernova \\
SNR & Supernova remnant \\
WD & White dwarf \\
XDINS & X-ray dim isolated neutron star \\

\end{tabular}}

\appendixtitles{no} 




\reftitle{References}


\externalbibliography{yes}
\bibliography{references}

\end{paracol}
\end{document}